\documentclass[twocolumn,showpacs,preprintnumbers,amsmath,amssymb]{revtex4}

\usepackage{graphicx}
\usepackage{dcolumn}
\usepackage{bm}

\begin{document}

\preprint{APS/123-QED}

\title{Hyperfine anomaly in Be isotopes in the cluster model and the neutron
spatial distribution.}

\author{Y. L. Parfenova}
 \email{Yulia.Parfenova@ulb.ac.be}
 \altaffiliation[Also at ]{ Skobeltsyn Institute of Nuclear Physics, Moscow State
University, 119992 Moscow, Russia}
\author{Ch. Leclercq-Willain}%
 \email{cwillain@ulb.ac.be}
\affiliation{%
Physique Nucl\'eaire Th\'eorique et 
 Physique Math\'ematique, CP229, Universit\'e Libre de Bruxelles
 B 1050 Brussels, Belgium.\\
}%


\date{\today}

\begin{abstract}
The study of the hyperfine anomaly of neutron rich nuclei, 
in particular, neutron halo nuclei, can give a very specific 
and unique way to measure their neutron distribution and 
confirm a halo structure. The hyperfine structure anomaly 
in Be$^+$ ions is calculated  with a realistic electronic
wave function, obtained as a solution of the Dirac equation.
In the calculations, the Coulomb potential modified by the charge 
distribution of the clustered nucleus and three electrons in 
the $1s^22s$ configuration is used. 
The nuclear wave function is obtained in the core+nucleon model 
of $^{9,11}$Be. The aim of this study is to test whether the 
hyperfine structure anomaly reflects a halo structure in 
$^{11}$Be.
\end{abstract}

\maketitle

\section{Introduction \label{sec:s1}}

Exotic (halo) nuclei are currently a subject of intensive
experimental and theoretical studies. The interest in 
neutron-rich nuclei is to a large extent driven by 
experimental facilities and new experimental methods for 
studying the nuclear matter distribution. In particular, 
the ion trap method \cite{Wad97} and the NMR (Nuclear 
Magnetic Resonance) methods \cite{Iso93} allow the 
measurements of the hyperfine splitting of electronic
states in atoms with an accuracy of the order $10^{-6}$, 
that provides the possibility of hyperfine anomaly studies. 

The hyperfine splitting
is sensitive to the magnetic current in the nucleus, and the 
hyperfine structure (hfs) constants extracted from the 
experimental data are related to the matter distribution 
of the nucleus. Therefore, the measurements of the hfs 
anomaly in neutron-rich halo nuclei can give a unique 
way to investigate the neutron distribution and the 
cluster structure.

For the experimental values of the magnetic moment of the Be isotopes, we
refer to the measurements in Refs. \cite{Web76,Ita83,Gei99}. 
But still there are no experimental data on the hyperfine anomaly for these
nuclei.

A theoretical study has been performed by Fujita \textit{et al.} 
\cite{Fuj99} in order to
verify how the halo structure manifests itself in the hyperfine 
structure. They calculated the hfs anomaly for Be isotopes 
both in the core+neutron model and in the single particle model, 
where the magnetic moment is well reproduced. They found 
that the hfs anomaly for $^{11}$Be is large compared to that for
the $^{7,9}$Be isotopes, and that this was indicative of an extended 
neutron distribution and a halo structure in $^{11}$Be. 

However, the approach used by Fujita \textit{et al.} in Ref. 
\cite{Fuj99} suffered from a poor knowledge of the ground 
state wave function of the $^{9,11}$Be isotopes. The value 
of the hfs anomaly $\epsilon $ defined by the folding of the 
electronic and nuclear wave functions is rather sensitive to 
the spatial distribution of the valence neutron wave function 
and to the weights of the possible mixed states in the description 
of the ground state wave function. During the last decade new information on 
the $^{11}$Be ground state wave function became available from 
the cross section measurements in the $p$($^{11}$Be,$^{10}$Be)$d$ 
reaction and the weight of the 2s$_{1/2}$ state admixture was 
found at the value 0.84 \cite{For99} which is close to the 
theoretical estimation.

Besides this, the cluster (core+neutron) model while being 
rather good for the $^{11}$Be nucleus, might fail for $^{9}$Be, 
which has essentially a three-cluster ($\alpha +\alpha +n$)
structure \cite{Des02}. On the other hand, there is no strong 
evidence for the three-body structure of $^9$Be in its ground 
state. So this is another interesting problem which can also 
be investigated in hfs anomaly studies.

The $^{11}$Be $1/2^+$ ground state can be represented by an 
admixture of the 2s$_{1/2}$ and 1d$_{5/2}$ valence neutron 
states, $^{10}Be (0^{+})\otimes \nu(2s_{1/2})$ and $^{10}Be 
(2^{+})\otimes \nu(1d_{5/2})$. The weight of the $s$- state, 
related to the spectroscopic factor of this configurations 
obtained in both the shell model and the simple excited core 
cluster model, are found between 0.5 and 0.8 (for more 
information on the spectroscopic factors we refer to Refs. 
\cite{War92,Nun96,Aut70}). 
In our calculations we take the weights of the states $w(2s_{1/2}\otimes 0^{+})=0.72$ 
and $w(1d_{5/2}\otimes 2^{+})=0.28$, consistent with the experimental
data \cite{Aum00} (for more details, see \cite{Par00}).

Let us mention that in Ref. \cite{Fuj99} the hfs anomaly has
been calculated within the Bohr-Weisskopf approach, where the 
electronic wave function is found for the interior of the nucleus. 
It is not applicable to the $^{11}$Be nucleus, where a 
significant part (88\%) of the valence nucleon wave function 
is outside of the range of nuclear potential, indicating a 
halo \cite{Fed93}. In particular, the $^{10}$Be core root mean 
square (rms) radius 2.61 fm is rather small compared to the 
mean core-neutron distance (6 fm). Thus, in Ref. \cite{Fuj99}
a rather simplified electronic wave function is used which 
significantly differs from the direct numerical solution of 
the Dirac equation in the region $r < 10$ fm. Besides this, 
in calculations of the electronic wave function, the electron 
screening effects of the Coulomb potential were not taken into 
account, although this effect can also be essential in the
calculations of the hfs anomaly.

In the present paper, the relativistic electronic wave functions are
calculated for the extended nuclear charge density distribution defined by
the cluster structure of the nucleus \cite{deV87}. We assume that two 
electrons in the $1s$ state form the closed shell and are relatively unperturbed 
by the third
electron in the $2s$ state, and we define the wave function of this
third electron as a solution of the Dirac equation for a nuclear charge
potential screened by the $1s^2$ electronic closed shell 
\cite{Ros78}.

We compare our electronic wave function with that of Fujita 
\textit{et al.} \cite{Fuj99}. We analyze the hfs of $^{9,11}$Be 
with regard to the available experimental information using 
more realistic descriptions of both the nuclear and the 
electronic wave functions. The aim of this paper is to
answer the questions whether the hfs anomaly in these 
isotopes reflects the halo structure, and whether the hfs 
anomaly for $^{11}$Be isotope is larger than for $^{9}$Be.

\section{Magnetic hyperfine structure \label{HFSsec}}

The magnetic hyperfine interaction Hamiltonian is defined by 
\begin{equation}
\mathcal{H}=-\int \mathbf{J}(\mathbf{r})\cdot \mathbf{A}(\mathbf{r})\,d^{3}r. 
\end{equation}
Here, $\mathbf{J}$ is the nuclear current density, $\mathbf{A}$ is the
vector potential created by the atomic electrons 
\begin{equation}
\mathbf{A}(\mathbf{r})= \frac{1}{c} \int \frac{\mathbf{j}(\mathbf{r}^{\prime
})}{|\mathbf{r}-\mathbf{r}^{\prime }|}d^{3}r^{\prime } ,  
\end{equation}
where $\mathbf{j}(\mathbf{r}^{\prime })$ denotes the electron current
density operator 
\begin{equation}
\mathbf{j}(\mathbf{r}^{\prime })= - \; e \;\mathbf{\alpha}_e \delta(\mathbf{r%
} - \mathbf{r}^{\prime})  
\end{equation}
and $\mathbf{\alpha}_e$ is the Dirac matrix for relativistic electrons.

For a $N$-electron system, the hyperfine interaction Hamiltonian can be
defined as 
\begin{equation}
\mathcal{H} = - \frac{1}{c} \int \int \frac{\mathbf{J}(r) \cdot \mathbf{j}(\mathbf{%
r}^{\prime })} {|\mathbf{r}-\mathbf{r}^{\prime }|} \, d^{3}r\;
d^{3}r^{\prime}  .  \label{ham1}
\end{equation}

With the Neumann expansion of 
$\frac{1}{|\mathbf{r}-\mathbf{r}^{\prime }|}$
in equation (\ref{ham1}), the interaction Hamiltonian takes the form: 
\begin{eqnarray}
\mathcal{H} &=& - \frac{1}{c} \int \int \mathbf{J}(\mathbf{r}) 
\cdot \mathbf{j}(\mathbf{r}^{\prime }) 
[\mathbf{T}(\mathbf{r}_<) \cdot \mathbf{U}(\mathbf{r}_>)]_{(%
\mathbf{r}\mathbf{r}^{\prime})}d^3r d^3r^{\prime}  \nonumber \\
&=& + \frac{e}{c} \;\int \mathbf{J}(\mathbf{r})\;\mathbf{\alpha}_e [%
\mathbf{T}(\mathbf{r}_<) \cdot \mathbf{U}(\mathbf{r}_>)]_{(\mathbf{r}\mathbf{r}%
_e)}\;d^{3}r   ,   
\end{eqnarray}
where 
\begin{eqnarray}
\mathbf{T}(\mathbf{r}_<) \cdot \mathbf{U}(\mathbf{r}_>) &=& \sum\limits_{\lambda}
\;\sum\limits_{\nu} \; T_{\lambda \nu}(\mathbf{r}_<) \; U_{\lambda \nu }^*(\mathbf{%
r}_>)  \\
&=& \sum\limits_{\lambda} \sum\limits_{\nu} (-)^\nu \frac{ r_<^\lambda}{r_>^{\lambda+1}} 
C_\nu^\lambda (\hat{r}%
_<)  C_{-\nu}^\lambda(\hat{r}_>)    \nonumber
\end{eqnarray}
and 
\[
C_\nu^\lambda(\hat{r})= \sqrt{\frac{4 \pi}{2 \lambda + 1}}\;Y_{\lambda \nu}(%
\hat{r}) 
\]
with
$r_<$ and $r_>$ being the smallest and largest value of the nuclear ($r$) or
electronic ($r_e$) coordinates. The $\lambda=1$ term is the magnetic dipole
interaction between the magnetic field generated by the electrons and the
nuclear magnetic dipole moment due to the extended nuclear matter
distribution. The $\lambda=2$ term is the electric quadrupole interaction
between the electric field gradient from the electrons and the non-spherical
charge distribution of the nucleus.

The hyperfine interaction couples the electronic angular momentum $\mathbf{J}$
and the nuclear one $\mathbf{I}$ to an hyperfine momentum $\mathbf{F} = \mathbf{J}
+ \mathbf{I}$. The magnetic hyperfine splitting energy $W$ for a state $\mid I
J F M_F=F >$ is defined as the matrix element of the Hamiltonian $\mathcal{H}
$,

\begin{widetext} 
\begin{eqnarray}
W_{(I J)F F } &=& < I J F F \mid \mathcal{H}\mid I J F F >    \\
&=& \sum\limits_{m,m\prime} < I J F F \mid I M , J m >     
 < I M , J m \mid \mathcal{H}\mid I M^\prime , J m^\prime >   
 < I M^\prime , J m^\prime \mid I J F F >  \nonumber
\end{eqnarray}
with the matrix element
\begin{equation}
< I M , J m \mid \mathcal{H}\mid I M^\prime , J m^\prime >
=  -2 i \frac{e}{c}\;\sum\limits_{\lambda} < I M \mid  
 \int \mathbf{J}(\mathbf{r})\cdot \;( 
\mathbf{A}_\lambda^0(r) + \mathbf{A}_\lambda^c(r))d^3r\;\mid I M^\prime > 
,    
\end{equation}
where
\begin{equation}
\mathbf{A}_\lambda^0 (\mathbf{r})
 =  r^\lambda \sum\limits_{\nu} (-)^\nu C_\nu^\lambda(\hat{r}) 
\int\limits_{0}^{\infty} dr_e \frac{r_e^2}{(r_e)^{\lambda+1}} \; F^{\kappa
J}(r_e)\;G^{\kappa J}(r_e)  \nonumber   \\
 < J m \mid C_{-\nu}^\lambda(\hat{r_e}) \mathbf{\sigma}_e \mid J m^\prime>  
\end{equation}
is the expression for a point nucleus, and
\[
\mathbf{A}_\lambda^c (\mathbf{r})= r \;\sum\limits_{\nu} (-)^\nu \;C_\nu^\lambda(%
\hat{r})\;\int\limits_{0}^{r} dr_e \; F^{\kappa J}(r_e)\;G^{\kappa J}(r_e) \;(\frac{%
r_e^{\lambda+2}}{r^{\lambda+2}}\;-\;\frac{r^{\lambda-1}}{r_e^{\lambda-1}}) <
J m \mid C_{-\nu}^\lambda(\hat{r}_e)\;\mathbf{\sigma }_e\;\mid J m^\prime> 
\]
\end{widetext} 
is the correction term due to the finite extension of the nuclear 
density. 

The functions $F^{\kappa J}$, $G^{\kappa J}$ are the radial parts
of the large and small components of the Dirac wave function of the
electron, with the quantum number 
${\kappa}=\pm (J+\frac{1}{2})$ for $J=L \mp \frac{1}{2}$
and the orbital angular momentum $L$.

In the dipole approach ($\lambda=1$) the expression for 
$W_{(I J)F F}$ reduces to
\begin{eqnarray}
W_{(I J)F F } &=& <IJFF|\mathcal{H}|IJFF>    \\
&=& \frac{1}{2}[F(F+1)-I(I+1)-J(J+1)] a_{I} \nonumber
\end{eqnarray}
where $a_I$ is defined by 
\begin{equation}
a_{I} = -\frac{2e \kappa \mu _{N}}{IJ(J+1)} < I I \mid \sum\limits_{i=1}^{A}
(M_Z^{l}(\mathbf{r}_i) + M_Z^{s}(\mathbf{r}_i))\mid I I >    \nonumber
\end{equation}
with the $Z$ components of the angular (or spin) magnetic moments $\mathbf{M}%
^{l(s)}(\mathbf{r}_i)$. Here, the summation runs over all nucleons.

Taking into account the density expansion of the nucleus, these 
moments are 
\begin{eqnarray}
\mathbf{M}^{l}(\mathbf{r}_{i}) &=&g_{l}^{i}\mathbf{l}_{i} [\int \limits_{r_{i}}^{%
\infty }F^{\kappa J} G^{\kappa J}\;dr\;+\;\int \limits_{0}^{r_{i}}F^{\kappa
J} G^{\kappa J}\;(\frac{r}{r_{i}})^{3} dr] ,    \nonumber \\
\mathbf{M}^{s}(\mathbf{r}_{i}) &=&g_{s}^{i}[\mathbf{s}_{i} \int \limits_{r_{i}}^{%
\infty }F^{\kappa J} G^{\kappa J}dr + \mathbf{D}%
_{i} \int \limits_{0}^{r_{i}}F^{\kappa J} G^{\kappa J}(\frac{r}{r_{i}}%
)^{3} dr] ,     \nonumber
\end{eqnarray}
where $\mathbf{D}_{i}={-\sqrt{\frac{5}{2}}\;[\mathbf{s}^{1}\otimes C^{2}(%
\hat{r}_{i})]}^{1}$.

Thus $a_I$ can be expressed through the hfs constant for a point 
nucleus $a_I^{(0)}$ as 
\begin{equation}
a_I \cong a_I^{(0)} (1+\epsilon+\delta)  , \label{eq10}
\end{equation}
where $\epsilon$ is defined as the hfs anomaly in the 
Bohr-Weisskopf effect 
and $\delta$ is the Breit-Rosenthal-Crawford-Shawlow ('BRCS') 
correction \cite{Ros72}. 

The hfs constant for the point nucleus is
\begin{equation}
a_I^{(0)}= -\frac{2 e \kappa \;\mu _{N} \mu}{IJ(J+1)} 
\int\limits_{0}^{\infty} F_0^{\kappa J}(r) G_0^{\kappa J}(r)dr ,   
\end{equation}
where $\mu =< I I \mid \sum\limits%
_{i=1}^{A} g_s^i s_i + g_l^i  l_i \mid I I >$ 
defines the magnetic moment of the point nucleus in nuclear
magneton units $\mu _{N}$. The functions $F_0^{\kappa J}(r)\;,\;G_0^{\kappa
J}(r)$ are the radial parts of the large and small components of the Dirac
wave function for the electron in the point nucleus approximation.

So, the hfs anomaly is defined by
\begin{eqnarray}
\epsilon & = & - \frac{b}{\mu}
\sum\limits_{i=1}^A \left\{ <II| (g_s^{(i)} s_i + g_{l}^{(i)} l_i) K^a
(r_i)|II> \right. \nonumber \\
&-& \left. <II| (g_l^{(i)} l_i + D_i) K^b(r_i) \mid II> \right\} .      
\end{eqnarray}
Here, $b=[\int_0^{\infty} F_0^{\kappa J}G_0^{\kappa J}d r]^{-1}$
is a constant obtained from (\ref{eq10}) and
\begin{eqnarray}
K^a(r_i) &=& \int\limits_{0}^{r_i}F^{\kappa J}G^{\kappa J}d r          \label{elwfinta} \\
K^b(r_i) &=& \int\limits_{0}^{r_i}F^{\kappa J}G^{\kappa J} (\frac{r}{r_i})^3 d r. \label{elwfintb}
\end{eqnarray}

The 'BRCS' correction is $\delta=1-b K^a(\infty)$ \cite{Ros72}.

Let us assume the two-cluster nuclear wave function defined as a superposition
of different configurations $\Phi^I_{L,S} = [\Phi_{l,s_c}^{J_c} \otimes
\varphi_{l,s}^{j}]^{I}$, associated to the coupling of a core
state $\Phi_{l,s_c}^{J_c}$ and the valence particle wave function $%
\varphi_{l,s}^{j}$. $J_c$, $s_c$ and $j$, $s$ are the total angular momentum and
spin of the core fragment and valence nucleon, $l$ is the orbital angular
momentum of their relative motion. For each configuration, the 
contribution to the hfs anomaly is given by 
\begin{eqnarray}
\epsilon&=&- \frac{b}{\mu} 
\sum\limits_{i=1,2}\left[<II| (g_s^{(i)} s_i + g_l^{(i)}\frac{m_{3-i}}{M} l)
K^a (\frac{m_{3-i}}{M} R) |II> \right.   \nonumber \\
&-& \left.<II|( g_{l}^{(i)}\frac{m_{3-i}}{M} l + D_i )K^b(\frac{m_{3-i}}{M} R)|II>
\right]    \label{anom}
\end{eqnarray}
where indices correspond to the core ($i=1$) and valence nucleon ($i=2$),
$M = m_1 + m_2$ is the mass of the whole system; 
$r$ and $R$ define, respectively, the electronic
and relative radial coordinates of the two nuclear fragments.
$g_l^{(i)}$
and $g_s^{(i)}$ are the gyromagnetic ratios of the $i$-th fragment orbital
motion and spin, respectively.

\section{Electronic wave functions \label{sec:s3}}

In the experiment \cite{Wad97}, the hyperfine splitting of the electronic
levels ($F$) is measured for\ the ground state of Be$^{+}$ ions. The
electronic ground state of the Be$^{+}$ ion is represented by the 
$1s^{2}2s$ configuration, and the hyperfine splitting of the
states $F=2$ and $F=1$ is found at about 1.256 GHz.

In the present paper, we calculate the hfs anomaly of the Be 
isotopes using the electronic wave functions 
$F^{\kappa J}= \frac{f(r)}{r}$ and 
$G^{\kappa J}=\frac{g(r)}{r}$ obtained as solutions of the 
Dirac equation \cite{Dav65} taking into account the $(1s^{2})$ 
electron screening and the extended density of the nuclear charge 
distribution. The electronic wave functions in the region 
$r\leq 22$ fm are obtained for the Coulomb potential defined by
the expression 
\begin{equation}
V(r)= 4\pi\frac{Z\alpha \hbar c}{r} \left\{
\int\limits_{0}^{r}\rho \; x^{2}dx+r\int\limits_{r}^{\infty
}\rho \; x \; dx\right\}  ,  \label{Vnucl}  
\end{equation}
where $\rho(x)$ is the charge distribution of the nucleus. This distribution
is obtained in the cluster model of the Be isotopes using the ground state
nuclear wave functions. To calculate the electronic wave functions we use 
the numerical
methods suggested in Ref. \cite{Ros61}.

The asymptotic electronic wave functions ($r\geq 22$ fm) are found in the
form 
\begin{eqnarray}
f(r)&=& e^{-D r}\sum \limits_{\nu}^{N}a_{\nu} r^{s+\nu} , \nonumber \\
g(r)&=&\frac{s-k}{\alpha Z_{eff} }e^{-D r}\sum \limits_{\nu}^{N} b_{\nu} r^{s+\nu}  ,  \nonumber 
\end{eqnarray}
where $A =\frac{1}{\hbar c}(E_{N} + mc^{2})$, $B =\frac{1}{\hbar c}(-E_{N} +
mc^{2})$, $D =\sqrt{A B}$, $s =\sqrt{k^{2}-(\alpha Z_{eff})^{2}}$ and the 
$2s$ electronic energy $E_N$ given by
\begin{equation}
E_{N}= mc^2 \left[ 1+\left( \frac{Z_{eff}\alpha }{N+\sqrt{%
k^{2}-(Z_{eff}\alpha )^{2}}}\right) ^{2}\right] ^{-1/2}  \label{Energy}
\end{equation}
with $N=0,1,2...$, and $k=\pm 1,\pm 2,\pm 3...$, are found with 
the effective charge $Z_{eff}$ depending on radius $r$. 
$a_{\nu}$ and $b_{\nu}$ are expansion coefficients (see Ref. \cite{Dav65}).

In the case of a Be$^{+}$ ion, there are three electrons orbiting the nucleus.
We can make a simplification by considering the $2s$ electron in the
Coulomb potential of the nucleus screened by the closed electronic shell $%
1s^{2}$. Thus, the asymptotic electron wave functions are defined for
the screened Coulomb potential written in the form: 
\begin{eqnarray}
V(r)&=&\frac{Z\alpha \hbar c}{r}\left[ 1-2\pi \left\{
\int\limits_{0}^{r}\rho _{el} x^{2}dx+r\int\limits_{r}^{\infty }\rho
_{el} xdx\right\} \right.  \nonumber \\
&+& \left. \frac{1}{2}(81/8\pi )^{1/3}\rho _{el}^{1/3}(r)\right] , 
\label{Vpot}
\end{eqnarray}
where $\rho _{el}(x)$, is the electron density distribution of the two $1s$
electrons in the closed shell. This potential can be approximated by 
\begin{equation}
V(r)=\frac{Z\alpha \hbar c}{r}\frac{1}{2}(e^{-\varkappa r}+1) , 
\end{equation}
where $\varkappa $ is fitted to reproduce the initial potential. So, the
effective charge for the $2s$ electron in the Be$^+$ ion is \\
\begin{equation}
Z_{eff}=\frac{Z}{2}(e^{-\varkappa r}+1)
\end{equation}
with $\varkappa =0.000068$ fm$^{-1}$.

Finally, the electronic wave functions calculated in the interior of the
nucleus are matched to those obtained
with the screened Coulomb potential.

\section{Results and discussion \label{sec:s4}}

The hfs anomaly $\epsilon $ for the Be$^{+}$ ion is obtained from the
relation $a_{I}= a_{I}^{(0)}(1+\epsilon +\delta)$ where $a_{I}^{(0)}$ is the
Fermi-contact parameter found for a point nucleus \cite{Bel63}. In the
core+nucleon model of the Be isotopes the values $a_I$ and $\epsilon$ are
determined by the valence nucleon wave function (that, in general, is an
admixture of different single particle states) and the electronic wave 
function depending on the charge distribution of the nucleus. Hence, the
amount of clustering or the existence of a "halo" in the Be isotopes nuclei
can be analyzed through the "hfs" anomaly evaluation.

\subsection{$^{11}$Be}

We consider the $^{11}$Be nucleus as a two-body system composed of a $^{10}$%
Be core nucleus in different states, 0$^+$ and 2$^+$, and a valence neutron. 
The $^{11}$Be
ground state is described by a superposition of the neutron states, $%
2s_{1/2}$ and $1d_{5/2}$, as
\begin{eqnarray}
\left| ^{11}{\rm Be} \left( \frac{1}{2}^{+}\right) \right\rangle &=& \beta \left|
[^{10}Be \left( 0^{+}\right) \otimes n_{s_{1/2}}]_{1/2^{+}}\right\rangle   \label{WF}  \\
&+& \lambda \left| [^{10}Be \left( 2^{+}\right) \otimes
n_{d_{5/2}}]_{1/2^{+}}\right\rangle       \nonumber
\end{eqnarray}
with the weights $w_s=\beta^2$ and $w_d=\lambda^2$ of the $s$- and $d$-waves
obtained in \cite{Aum00}, which allow a good description of the $^{11}$Be 
interaction
and break-up reaction cross sections at intermediate and high energies 
(see Ref. \cite{Par00}).

In our approach, the magnetic moment of $^{11}$Be in the ground 
state is 
\begin{eqnarray}
\mu&=&w_s \mu_s+w_d \sum_{m_{s},m_{l}}\sum_{m_{j},M_{S}}
\left(C_{sm_{s}lm_{l}}^{jm_{j}}C_{SM_{S}jm_{j}}^{\frac{1}{2}\frac{1}{2}}\right)^{2} \nonumber \\
& \times & [\frac{\mu_d}{1/2}m_{s}+\frac{\mu(2^{+})}{2}M_{S} + \mu_c \frac{m_l}{2}]
   \\
&=&w_s \mu_s +\frac{7}{15} w_d \mu_d+ 
 \frac{7}{15} w_d \mu_c -\frac{1}{3} w_d \mu (2^{+}), \nonumber
\end{eqnarray}
where $(s, m_{s})$, $(l, m_{l})$, and $(j , m_{j})$ are
the valence neutron spin, orbital angular momentum and total momentum,
and their projections. $S$, $M_{S}$ are the spin
of the $^{10}$Be core nucleus and its projection. The magnetic moment of the 
$^{10}$Be core fragment related to its orbital motion is $\mu_c = \frac
{Z}{A(A-1)} l$, $A=11$. Here, $\mu_s$, $\mu_d$, and $\mu (2^{+})$ denote the 
magnetic moment of the neutron
in the $s_{1/2}$ and $d_{5/2}$ states and the $^{10}$Be core in the 2$%
^{+} $ state. We use the values $\mu_s=\mu_d=\frac{1%
}{2}g_{s}^{(n)}=-1.9135 \mu_N$. 
The magnetic moment of $^{10}$Be in the excited state $%
I^{\pi}=2^+$ has been calculated in Ref. \cite{Suz95} in the shell model,
and has been found to be $\mu (2^{+})=1.787\mu_{N}$. 

The ground state wave function (\ref{WF}) gives the $^{11}$Be magnetic 
moment value $-1.784 \mu_N $ that
is larger than the experimental value $\mu_I=-1.6816(8) \mu_N $ \cite{Gei99}.
This problem with the magnetic moment description was discussed, in
particular, in \cite{Ots93}, and the weights of the $s$- and $d$- waves
have been found $w_s$=0.55, and $w_d$=0.45. In our simplified model, 
we reproduce the
experimental value of the $^{11}$Be magnetic moment with 
$w_s$=0.5, and $w_d$=0.5, 
which are the same as those used in calculations \cite{Fuj99}. 
This example gives a measure of the sensitivity 
and the precision of the calculations.

The valence neutron wave function for each partial state in
$^{11}$Be is obtained as a solution of the
Schr\"{o}dinger equation with the same Woods-Saxon potential
parameters as in Ref. \cite{Par00}, and reproduce the neutron separation 
energy $B_n$. These parameters are given in Table \ref{tab1}.

\begin{table}[tbp]
\caption{\label{tab1} Core and valence neutron parameters used for the 
calculation of the
core-nucleon wave function and core density parameters. $E_{x}$ is the
excitation energy of the core state, $w_j$ are the weights 
of the different configurations , $V_{0}$, $R_{0}$ are the parameters of
the Woods-Saxon potential with diffuseness parameter $a_{0}= 0.5$ fm, $B_n$
is the neutron separation energy.}
\begin{ruledtabular}
\begin{tabular}{ccccccc}
$E_{x}$ & core & neutron &  & \multicolumn{3}{c}{Woods-Saxon
potential $^{b}$}  \\ \hline
      & $J_c^{\pi}$ & $nl_{j}$ & $w_j$   & $V_{0}$  & $R_{0}$ & $B_{n}$  \\
(MeV) &         &            &         & (MeV)    & (fm)    & (MeV)   \\ \hline
0     & $0^{+}$ & $2s_{1/2}$ & $0.719$ & $-70.83$ & $2.483$ & 0.504  \\
3.368 & $2^{+}$ & $1d_{5/2}$ & $0.281$ & $-89.22$ & $2.483$ & 3.872  \\ 
\end{tabular}
\end{ruledtabular}
\\ $^{b}$ The $^{10}$Be charge rms radius is related to the matter
radius $r_{m}$ as rms=$\sqrt{r_{m}^2+0.64}$, where $r_{m}$ is taken
from Ref. \cite{Oza01} \hphantom{1}
\end{table}

The $^{10}$Be core density is parametrized using the harmonic oscillator
model \cite{deV87} 
\begin{equation}
\rho(r)= \rho _{0} (1 + \gamma (r/a)^2) exp(-(r/a)^2),   \label{density} 
\end{equation}
where $\rho _{0}$ is a normalization factor, $a=1.856$ fm, 
and $\gamma =0.610$.

The charge density distribution of the $^{11}$Be nucleus is defined as 
\begin{equation}
\rho (r)=\rho _{0}\int d^{3}\mathbf{x}\,\,\left|\Phi _{L}^{2}(x)\right|\rho
_{c}(|\mathbf{r}-\frac{1}{11}\mathbf{x}|) ,  \nonumber
\end{equation}
where $\rho _{0}$ is a normalization factor, $\rho_c$ is the core
density and $\Phi _{L}(x)$ defines the radial part of the wave function for
the valence neutron. The charge density distribution in (\ref{Vnucl}) and
the electronic density distribution in (\ref{Vpot}) give the Coulomb
potential entering the Dirac equation for the wave function
of the outer $2s$ electron.

With the weights from Table \ref{tab1} and parameters mentioned 
above we obtain the value of the hfs anomaly for the 
$^{11}$Be$^{+}$ ions as
\begin{eqnarray}
\epsilon &=&-\frac{b}{\mu_I} 
\left[ \beta^2 \mu_{s} 
\{ K_{0}^a(\frac{M_c}{M})- (1-\Sigma_{ \frac{1}{2}}) K_{0}^b(\frac{M_c}{M}) \} \right.   \nonumber \\
&+& \left. \frac{7}{15}\lambda^2 \mu_{d}
\{ K_{2}^a(\frac{M_c}{M})- (1-\Sigma_{\frac{5}{2}}) K_{2}^b(\frac{M_c}{M}) \}\right] \label{anom11} \\ 
&+&\frac{1}{3} \frac{b}{\mu_I} 
\lambda^2 \{\mu(2^+) \tilde{K_{1}^a}(\frac{m}{M})+ \Sigma(2^+)\tilde{K_{1}^b}
(\frac{m}{M}) \} \nonumber \\
&-& \frac{7}{15} \lambda^2 \mu_c
\{ K_{2}^a(\frac{M_c}{M})- K_{2}^b(\frac{M_c}{M}) \}  ,  \nonumber 
\end{eqnarray}
where
$$
\Sigma_{j}=\pm \frac{3}{4} \frac{j+1/2}{j+1}
$$
for $j = l \pm \frac{1}{2}$. The value of $\Sigma(2^+)$ is taken as
$\Sigma(2^+)=-1.034$ (see Ref. \cite{Fuj99}). Here, $m$ and $M_c$ are the mass of the 
neutron and of the  core $^{10}$Be, respectively. 
The ratio $\frac{M_c}{M}$ ($\frac{m}{M}$) takes into account the 
motion of the core (valence neutron) relative to the $^{11}$Be
center of mass (cm).

The terms $K_{l}^{a}$ and $K_{l}^{b}$ are expressed as 
\begin{eqnarray}
K_{l}^{a}(m_i) &=&\int\limits_{0}^{\infty }|\Phi
_{l}(R)|^{2}R^{2}dR K^{a}(m_i R)             \label{wffolda} \\               
K_{l}^{b}(m_i) &=&\int\limits_{0}^{\infty }|\Phi
_{l}(R)|^{2}R^{2}dR K ^{b} (m_i R)           \label{wffoldb}
\end{eqnarray}
where $K^{a}$ and $K^{b}$ are defined in (\ref{elwfinta}) and (\ref{elwfintb})
and $\Phi_{l}(R)$ is the radial part of the valence nucleon wave function,
for the state with the orbital angular momentum $l$. 

Notice, that the terms $\tilde{K_{1}^a}$ and $\tilde{K_{1}^b}$ 
in (\ref{anom11}) are calculated
for the $p$ - wave nucleons composing the excited $^{10}$Be nucleus 
and contributing to the magnetic
moment of the excited 2$^+$ state. In these calculations we used the folding 
of the electronic part $K^{a(b)}$ with
the $^{10}$Be density distribution, assuming that the spatial distribution
of the $p$ - wave nucleons is similar to the charge density distribution in 
$^{10}$Be.

Table \ref{tab2} shows the values of the hfs anomaly
calculated without screening effect ($\epsilon^{ns}$), or 
that for a homogeneously charged sphere 
with the rms radius 2.61 fm ($\epsilon^{hs}$),
and the value $\epsilon$ obtained by including all the effects. 
The contribution of the terms 
$\tilde{K_{1}^a}$ and $\tilde{K_{1}^b}$ in (\ref{anom11})
to the hfs anomaly are given in the row named "core".
In the Table, $r_n$ is the  mean distance of the valence 
neutron from the $^{11}$Be cm.

\begin{table}[tbp]
\caption{\label{tab2} The hfs anomaly in $^{11}$Be$^+$ calculated 
for different states of the $^{10}$Be core.  $B_{n}$ is the 
neutron separation energy, $r_n$ is the neutron distance, 
$\protect\epsilon^{ns}$ is the hfs anomaly value obtained
without consideration of the screening, $\protect\epsilon^{hs}$ 
is calculated for a homogeneous spherical charge distribution, 
and $\protect\epsilon$ is obtained including all the effects. 
In the last two rows the weighted
sums of the values are given.}
\begin{ruledtabular}
\begin{tabular}{cccccccc}
State & $B_{n}$ & $r_n$ & $\epsilon ^{ns}$ & 
$\epsilon ^{hs}$ & $\epsilon $ & $r_n$ $^{c}$  
& $\epsilon$ $^{c}$  \\ 
$nl_j$      & (MeV) & (fm)  &  (\%)   &  (\%)   &  (\%)   & (fm)  &  (\%)  \\ \hline
$2s_{1/2}$  & 0.504 & 6.056 & -0.0558 & -0.0694 & -0.0719 & 6.165 & -0.120 \\ 
$1d_{5/2}$  & 3.872 & 2.634 & -0.0090 & -0.0103 & -0.0112 & 3.551 & -0.0233 \\
core        &       & 2.623 & -0.0006 & -0.0007 & -0.0008 & 2.588 &        \\   \hline
\multicolumn{3}{l}{Total, $w_s=0.72$}       
                            & -0.0433 & -0.0534 & -0.0556 &       &        \\ \hline
\multicolumn{3}{l}{Total, $w_s=0.5$}        
                            & -0.0335 & -0.0410 & -0.0430 &       & -0.0717 \\ 
\end{tabular}
\end{ruledtabular}
\\ $^{c}$ cited from Ref. \cite{Fuj99} \hphantom{1111111111111111111111111111111111}
\end{table}

One can see from the Table, that the hfs anomaly is sensitive 
to the screening effect (about 25\%). The total value weakly 
depends on the shape of the charge density distribution 
(about 3 \%), but the effect is different for distinct 
partial states, and is more pronounced for the $d$ - wave 
neutron (about 8\%).

These values of $\epsilon$ can be compared with those 
from \cite{Fuj99}, $\epsilon (2s_{1/2}) = -0.120 \%$ and 
$\epsilon (1d_{5/2}) = -0.0233 \%$. The values from \cite{Fuj99} 
are obtained without screening effect in
a simplified model of the nuclear charge distribution, 
approximated by a homogeneously charged sphere with the 
radius $1.2 A^{\frac{1}{3}} = 2.669$ fm.

The value of the 'BRCS' correction is $\delta=0.0476$\% that is
comparable with $\epsilon$.

The hfs anomaly value for each partial wave function 
strongly correlates with
the neutron separation energies $B_n$ and $r_n$.
Thus, the hfs anomaly value, obtained for the $s$-wave neutron 
is larger than that for the $d$-wave by an order of magnitude. 
Hence, the value of the
hfs anomaly is very sensitive to the weights of the partial waves.
For example, with the weights from Ref. \cite{Fuj99} $w_s=w_d=0.5$
we get the value $\epsilon=$ -0.0430\% which is essentially smaller than that
obtained in Ref. \cite{Fuj99}, $\epsilon=$ -0.0717\%.
The contribution of the core fragment is also small.

Thus, the precise measurement of the hfs anomaly might shed light upon 
the spectroscopic factors of the $s$- and $d$- states and the origin 
of the $\frac{1}{2}^+$ state of $^{11}$Be.

The halo nucleus charge distribution might 
significantly deviate from a spherical distribution,
defining the radial dependence of the electronic wave function.
At the same time, the $^{10}$Be$ - n$ mean distance is $6.7$ fm, 
so that $r_n \simeq 6$ fm, a value exceeding a few times
the $^{10}$Be core rms radius. Thus a significant part
(88\%) of the valence nucleon
wave function is outside of the range of the nuclear potential. 
Hence, the
result of the folding of the electronic and nucleon wave functions
essentially depends on the radial behavior of the wave functions.

\begin{figure}
\includegraphics{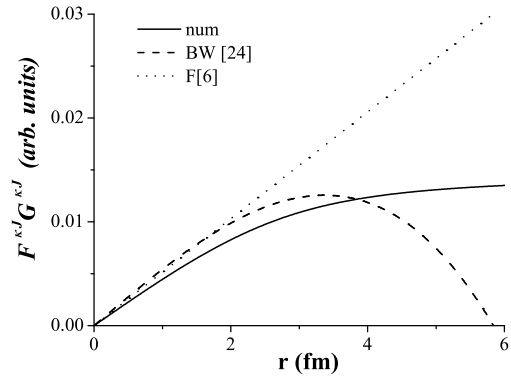}
\caption{\label{figeps} Electronic wave functions 
($J=\frac{1}{2}$, $\kappa=-1$) inside the 
$^{11}$Be nucleus, obtained in the Bohr-Weisskopf approach 
\cite{Boh50} (BW - dashed line), the simplified approach of 
Fujita \textit{et al.} \cite{Fuj99} (F- dotted line), and the 
numerical solution of the Dirac equation (num - solid line).}
\end{figure}

Fig \ref{figeps} shows the radial
dependence of the electronic wave functions inside the $^{11}$Be 
nucleus, calculated in the Bohr-Weisskopf approach \cite{Boh50}, 
the simplified approach \cite{Fuj99}, and that obtained here as 
a numerical solution of the Dirac equation. One can see that the 
calculated wave functions are close to each other at 
$r \leq 3$ fm and significantly different at higher radii. 
For $^{11}$Be, the folding of the electronic wave function with 
the wave function of the weakly bound valence nucleon
gives significantly different results. Clearly, with the wave
function of Fujita \textit{et al.} \cite{Fuj99}, 
the hfs anomaly is overestimated for weakly bound nuclei.

This overestimation, in particular, can be seen from comparison of the
hfs anomaly values for the $s$-wave valence neutron with close
$r_n$ values. The result of our calculations is smaller than
that from Ref. \cite{Fuj99}.
For the $d$-wave neutron, with different $r_n$ values
we obtain the hfs anomaly value twice as small as that from Ref. \cite{Fuj99}. 

Besides the difference in the electronic
wave function radial dependence, the assessment \cite{Boh50} 
of the integrals with the electronic wave function in 
(\ref{wffolda}) and (\ref{wffoldb}) as 
$ K_l^a(\frac{m_i}{M})-K_l^b(\frac{m_i}{M})=C K_l^a(\frac{m_i}{M})$ 
with $C=0.68$ is rather crude, and
the coefficient $C$ for different partial waves varies within 20\%.

To compare the calculated value of the hfs anomaly in $^{11}$Be with that 
for $^9$Be, as the next step
of our analysis, we perform similar calculations for the $^9$Be$^+$ ion.

\subsection{$^{9}$Be}

The hfs anomaly for $^{9}$Be can be calculated within the two-cluster model
of $^{9}$Be. This nucleus can be regarded as a system composed of a
$p_{3/2}$-wave neutron and a $^{8}$Be core in the ground (0$^{+}$) and 
excited (2$^{+}$) states. 

Note, that in general, the states 0$^{+}$, 2$^{+}$, and 4$^{+}$ are realized
in $^{8}$Be, but only 0$^{+}$ and 2$^{+}$ contribute to the hfs anomaly in
$^{9}$Be (corresponding to the valence neutron angular momentum $l=1$). The
contributions with $l=3$ are expected to be small.

The $^9$Be ground state is given by a 
superposition of the 0$^{+}$ and 2$^{+}$ states 
\begin{eqnarray}
\left| ^{9}{\rm Be} \left( \frac{3}{2}^{-}\right) \right\rangle 
&=&\beta \left|
[^{8}Be \left( 0^{+}\right) \otimes n_{p_{3/2}}]_{3/2^{-}}\right\rangle  \label{WF9}  \\
&+&\lambda \left| [^{8}Be \left( 2^{+}\right) \otimes
n_{p_{3/2}}]_{3/2^{-}}\right\rangle  .    \nonumber   
\end{eqnarray}

With this wave function, denoting $w_{0^+}=\beta^2$ and 
$w_{2^+}=\lambda^2$, the magnetic moment of $^{9}$Be in the 
core+cluster model is
\begin{equation}
\mu=w_{0^+} \{ \mu_p + \mu_c \}
+ w_{2^+} \{ \frac{1}{5}\mu_p + \frac{3}{5}\mu_c + \frac{1}{5}\mu (2^{+})\} .  
\end{equation}
Here, indices in $w_{J_c^\pi}$ correspond to the core states $J_c^\pi=0^+,2^+$.

The magnetic moment of the $^{8}$Be core related to its orbital 
motion is $\mu_c = \frac{Z}{A(A-1)} l$, $A=9$. 
Here, $\mu_p$, and $\mu (2^{+})$ 
denote the magnetic moment of the $p_{3/2}$ neutron
and the $^{8}$Be core in the 2$%
^{+} $ state, $\mu_p=-1.9135 \mu_N$ and $\mu (2^{+})=1$.

With the weights $w_{0^+}= 0.535$ and $w_{2^+}=0.465$ obtained
with the spectroscopic factors from Ref. \cite{Coh67} the magnetic 
moment is $\mu=-1.0687$ $\mu_N$.
The change of the parameters to $w_{0^+}=  0.579$
and $w_{2^+}=0.421$ allows one to reproduce the experimental value of
the $^{9}$Be magnetic moment $\mu_I=-1.1447 \mu_N$. This again gives a measure
of the sensitivity of the calculated value to the weights of the states.

In our approach, these two states of $^9$Be with the core 
in the ground and excited state are characterized by different neutron
separation energy $B_n$ listed in Table \ref{tab3}. For the case
with $^{9}$Be (0$^+$),
the valence neutron wave function is obtained as a solution of the
Schr\"{o}dinger equation for the Woods-Saxon potential 
$V_{0}=-$43.61 MeV, $a_{0}=0.5$ fm, and $R_{0}=2.46$ fm
giving $B_{n}=1.665$ MeV and
the $^{9}$Be charge rms radius 2.519 fm.

In the case of excited $^{8}$Be in the resonance state 2$^{+}$,
the parameters of the Woods-Saxon potential are taken
as $V_{0}=-$49.80 MeV, $a_{0}=0.5$ fm, and $R_{0}=2.49$ fm
to reproduce $B_{n}$ (see Table \ref{tab3}).

The $^{8}$Be core density is parametrized in the harmonic
oscillator model (\ref{density}), with the parameters
$a=1.749$ fm and $\gamma =0.619$ for the 0$^+$ state, and
$a=1.768$ fm and $\gamma =0.624$ for the 2$^+$ state, respectively.
The density parameters give the rms radius of the core equal to $R_0$
for each state.

The hfs anomaly for $^{9}$Be$^+$ ion is

\begin{eqnarray}
\epsilon &=&\frac{b}{\mu_I} 
\left[ (w_{0^+}+\frac{1}{5}w_{2^+}) \mu_{p} 
\{ K_{1}^a(\frac{M_c}{M})- (1-\Sigma_{\frac{3}{2}}) K_{1}^b(\frac{M_c}{M}) \}\right] \nonumber \\
&+&\frac{3}{5} \frac{b}{\mu_I} 
w_{2^+} \{\mu(2^+) \tilde{K_{1}^a}(\frac{m}{M})+ \Sigma(2^+)\tilde{K_{1}^b}(\frac{m}{M}) \} \label{anom9} \\
&+& (w_{0^+}+\frac{1}{5}w_{2^+}) \mu_c
\{ K_{1}^a(\frac{M_c}{M})- K_{1}^b(\frac{M_c}{M}) \}  ,  \nonumber
\nonumber
\end{eqnarray}
The values $K_{l}^a$ and $K_{l}^b$ are calculated with (\ref{wffolda})
and (\ref{wffoldb}). The values $\tilde{K_{1}^a}$ and $\tilde{K_{1}^b}$
are calculated with the density distribution of the core fragment in the
$^9$Be nucleus, taking into account the $^8$Be-neutron relative motion.

The calculations show that the contribution of the term
$\Sigma(2^+)\tilde{K_{1}^b}(\frac{m}{M})$ is small and we can neglect it.

In Table \ref{tab3} the hfs anomaly values are listed for
each $^8$Be state. The total value, obtained as the
weighted sum is given in the last row of the Table.
It is close to the values $\epsilon
=-0.0249$\% \cite{Fuj99}, $\epsilon =-0.0243$\% \cite{Yam00}.

The value of the 'BRCS' correction is $\delta=0.0451$\% that is
larger than $\epsilon$.

\begin{table}
\caption{\label{tab3} The hfs anomaly $\protect\epsilon$ 
in $^{9}$Be$^+$ calculated for different states
of the $^{8}$Be core here and in \cite{Fuj99}, 
$w_{J_c^{\pi}}$ are the weights of the different configurations.  
$B_{n}$ is the neutron separation energy, 
$r_n$ is the neutron distance. 
In the last row the weighted sum of the values is presented.}
\begin{ruledtabular}
\begin{tabular}{ccccccc}
Core & $w_{J_c^{\pi}}$ & $B_{n}$ & $r_n$ & $\epsilon $ & 
$r_n$ $^a$ & $\epsilon$ $^a$  \\ 
$J_c^{\pi}$ &  & (MeV) & (fm) & (\%) & (fm) & (\%)  \\ \hline
$0^{+}$  & 0.535 & 1.665 & 3.200 & -0.0440 &   2.569  & -0.0249  \\ 
$2^{+}$  & 0.465 & 4.705 & 2.630 & -0.0066 &        &        \\
core     & 0.535 &       & 2.490 &  0.0063 &        &      \\   \hline
Total    &        &       &       & -0.0236 &        & -0.0249  
\end{tabular}
\end{ruledtabular}
\\$^a$ cited from Ref. \cite{Fuj99} \hphantom{1111111111111111111111111111111111}
\end{table}

One can see that the result for $^9$Be is twice as small as that for
$^{11}$Be. This corresponds to the conclusion in Ref. \cite{Fuj99}, 
that the
value of the hfs anomaly reflects the extended neutron distribution
in $^{11}$Be and might indicate a neutron halo. At the 
same time, the difference in the hfs anomaly values is not as large
as in Ref. \cite{Fuj99}.

As with $^{11}$Be case, the $^9$Be hfs anomaly values for each 
partial wave function strongly correlate with the neutron distance 
$r_n$ from the Be cm.
In particular, the hfs anomaly obtained for the $p$-wave neutron for 
the $0^+$ state of the $^8$Be core is
larger than that for the $2^+$ state by an order of magnitude. The 
contribution of the core fragment is also small.


We should note, 
that we use a rather simplified model of $^{9}$Be which most likely
has a three-cluster structure ($\alpha+\alpha+n$). In the three-cluster
approach, there are partial states, contributing to the $^{9}$Be ground state 
wave function, which are not taken into account in the two-cluster
model, and which also contribute to the hfs anomaly value.

\section{Conclusion \label{sec:s5}}

In the present paper, we have calculated the hfs 
anomaly in the $^{9,11}$Be isotopes. The nuclear
wave functions have been calculated in the simplified core+neutron model.
In calculations of the realistic electronic wave function, the charge 
distributions of the clustered nucleus and electrons
in the $1s^22s$ configuration in Be$^+$ are taken into account.

It is found that the value of the hfs anomaly for $^{11}$Be 
($\epsilon=-0.0556$\% for $w_s$=0.72 and $\epsilon=-0.0410$\% 
for $w_s$=0.5) is larger than that for $^{9}$Be 
($\epsilon=-0.0235$\%), that agree with the conclusion 
of Fujita \textit{et al.} \cite{Fuj99}. 

For $^{9}$Be we reproduce the hfs anomaly value, obtained in \cite{Fuj99}, 
$\epsilon=-0.0249$\%, and for $^{11}$Be we get the values with different
weights $w_s$ of the $s$-wave, which are essentially
less than that in \cite{Fuj99}, $\epsilon=-0.0717$\%, obtained with the 
weight $w_s$=0.5.

In calculations of the hfs anomaly we did not consider
the three-cluster structure of $^{9}$Be. We did not take
into account the probable contribution of other states admixed in the ground
state wave function of $^{9}$Be, which also might influence the
calculated value of the anomaly. Thus, the difference in the hfs 
anomaly in the $^{9}$Be and $^{11}$Be isotopes might be even less.

Finally, we can conclude, that, on one hand, 
the hfs anomaly correlates with the neutron spatial distribution.
As the anomaly value is sensitive to the ground state wave function of
the nucleus, there is no unambiguous correlation 
between the hfs anomaly value and valence neutron distribution, 
and each nucleus requires a separate investigation.


\end{document}